\documentclass[aps,prl,twocolumn,groupedaddress]{revtex4}
\usepackage{latexsym} \usepackage{amsmath} \usepackage{dcolumn}
\usepackage{bm} \usepackage{graphicx} \usepackage{epsfig}

\usepackage[]{graphicx}
\usepackage[]{wrapfig}

\begin{document}

\title{Intermediate coupling model of cuprates: adding fluctuations to a
weak coupling model of pseudogap and superconductuctivity competition}

\author{R.S. Markiewicz}\affiliation{Physics Department, Northeastern University, Boston MA
02115, USA}
\author{Tanmoy Das}\affiliation{Physics Department, Northeastern University, Boston MA
02115, USA}
\author{Susmita Basak}\affiliation{Physics Department, Northeastern University, Boston MA
02115, USA}
\author{A. Bansil}\affiliation{Physics Department, Northeastern University, Boston MA
02115, USA}
\date{\today}
\begin{abstract}
We demonstrate that many features ascribed to strong correlation effects
in various spectroscopies of the cuprates are captured by a calculation of
the self-energy incorporating effects of spin and charge fluctuations. The
self energy is calculated over the full doping range from half filling to
the overdoped system. In the normal state, the spectral function reveals
four subbands: two widely split incoherent bands representing the remnant
of the two Hubbard bands, and two additional coherent, spin- and
charge-dressed in-gap bands split by a spin-density wave, which collapses
in the overdoped regime.  The resulting coherent subbands closely resemble
our earlier mean-field results. Here we present an overview of the
combined results of our mean-field calculations and the newer extensions
into the intermediate coupling regime.
\end{abstract}

\pacs{74.25.Dw,71.27.+a,74.40.-n,74.40.Kb,74.25.Jb} 
\maketitle\narrowtext

Over the past few years evidence has been mounting that correlation
effects in the cuprates are not as strong as previously thought and that
these materials may fall into an intermediate coupling regime. If so, the
transition to an insulator could be described in terms of Slater physics
by invoking a competing phase with long-range magnetic order, rather than
Mott physics requiring a disordered spin liquid phase with no double
occupancy. The most striking evidence in this direction perhaps is the
existence of quantum oscillations in underdoped cuprates associated
with small Fermi surface pockets.\cite{leyraud,hussey} Strictly, quantum oscillations have
only been observed in strong magnetic fields, so that it is possible that
the ordered phases are field-induced.\cite{yeh} Nevertheless, it is clear
that a phase with long-range order lies close in energy to the ground
state. Additional evidence comes from the recent calculation of the
magnetic phase diagram of the half-filled $t-t'-U$ model by Tocchio, {\it
et al}.\cite{TBPS} They found that the ground state is either non-magnetic
at small $U$ or possesses a long-range magnetic order, except for a small
pocket of spin-liquid phase at values of $U$ and $|t'|$ too large to be
relevant for the cuprates.  Further, the high-energy spectral weight
associated with the `upper Hubbard band' decreases with doping too fast to
be consistent with strong coupling (no double occupancy)
models.\cite{comanac,tanmoysw}

There have been a number of recent attempts to extend strong coupling
calculations into the intermediate coupling regime.  Yang, {\it et
al.}\cite{YRZ} have introduced a phenomenological self energy that is
similar to that of an ordered magnetic phase, but with a special
sensitivity to the magnetic zone boundary.  Paramekanti, {\it et
al}.\cite{paramekanti} have carried out variational resonant-valence bond
(RVB)-like calculations where the requirement of no double occupancy has
been relaxed.  However, given that the ground state is close to being
magnetically ordered, one could also approach intermediate coupling by
including effects of fluctuations in a weak coupling scheme, where the
lowest order (Hertree-Fock) solution already describes a long-range
ordered state. This article presents an overview of our ongoing work in
this direction.\cite{markiewater,tanmoysw}

We begin with a brief recapitulation of the mean-field calculations, which
by themselves can provide a good description of the doping dependence of
the low-energy coherent part of the electronic spectrum. The model is
nearly quantitative for electron-doped cuprates, where the competing
magnetic order is known to be $(\pi ,\pi)$ antiferromagnetic over the full
doping range. Such a model can describe many features of the hole-doped
cuprates as well, even though the competing order[s] are known to be more
complicated.\cite{tanmoy2gap,Gzm1} Within this model, for electron-doped
cuprates, the ground state at half-filling is an antiferromagnetic
insulator, doping simply shifts the Fermi level into the upper magnetic
band producing an electron pocket near $(\pi ,0)$, and the magnetization
decreases with doping until magnetism collapses in a quantum critical
point near optimal doping. This quantum critical phase transition in fact
involves two Fermi surface driven topological transitions\cite{tanmoysns},
the first near $x=0.14-0.15$ where the top of the lower magnetic band
crosses the Fermi level producing hole pockets near $(\pi /2,\pi /2)$, and
a second transition near $x=0.18$, where the hole and electron pockets
recombine into a single large Fermi surface. The model has been able to
describe angle-resolved photoemission (ARPES),\cite{nparm,kusko}
resonant inelastic x-ray scattering (RIXS),\cite{RIXS} and scanning tunelling miscoscopy (STM)\cite{tanmoytwogap} spectra, and the
unusual pairing symmetry transition with doping seen in penetration depth
measurements\cite{tanmoyprl}. Recently, quantum oscillations were observed in electron
doped cuprates at several dopings, showing a crossover from the hole
pocket at lower dopings to the large FS at the highest doping.\cite{Helm}
Furthermore, the areas of the FS pockets measured by quantum oscillations are well
predicted by the model for the electron doped case, while hole doped
cuprates remain controversial in this aspect.

Fluctuations modify the above picture in several ways. First, in
two-dimensional materials, critical fluctuations are well-known to
eliminate long-range order and drive the antiferromagnetic transition
temperature to zero in accord with the Mermin-Wagner theorem, so that over
a wide range of temperatures only a pseudogap remains. [The observed
N\'eel order is driven by small deviations from isotropic two-dimensional
magnetism.] These fluctuations can be accounted for in a self-consistent
renormalization scheme\cite{markieMW}, and are necessary to describe the
response of the system at higher temperatures. Fluctuations also modify
the low-temperature physics at higher energies, leading to the high-energy
kink or the waterfalls seen in ARPES\cite{ronning,graf,pan}, effects of
the ARPES matrix element
nothwithstanding\cite{lindroos,asensio,bansil,sahrakorpi,susmita}.
We have recently introduced the quasiparticle-GW (QP-GW)
scheme to account for these fluctuations.\cite{markiewater,tanmoysw} In
this way, we have been able to describe the doping dependence of the
optical spectra\cite{onose,onoseprb,uchida}, including both the
`Slater-like' collapse of the midinfrared peak with doping and the
`Mott-like' persistence of a high-energy peak into the overdoped regime.
The model also quantitatively accounts for the anomalous spectral weight
transfer to lower energies with doping in the cuprates.\cite{tanmoysw}

In a GW-scheme, the self-energy is calculated from a variant of the
lowest-order `sunset' diagram, a propagator dressed by the emission and
reabsorption of a bosonic operator,
\begin{eqnarray}\label{selfeng}
\Sigma({\bf k},\sigma,i\omega_n)=\frac{1}{2} \sum_{{\bf q},\sigma^{\prime}}\eta_{\sigma,\sigma^{\prime}}
\int_{-\infty}^{\infty}\frac{d\omega_p}{2\pi} \Gamma({\bf k},{\bf q},\omega_n,\omega_p)
%~~~~~~~~~~~~~~~~~~~~~
\nonumber\\
G({\bf k}+{\bf q},\sigma^{\prime},i\omega_n+\omega_p)
{\rm Im}[W^{\sigma\sigma^{\prime}}({\bf q},\omega_p)],
\end{eqnarray}
Here, $\sigma$ is the spin index. $W\sim U^2\chi$ denotes the interaction,
which involves the Hubbard $U$ and the susceptibility $\chi$ in the random
phase approximation (RPA)\cite{vignale}, and $\Gamma$ is a vertex
correction.\cite{foot0} $\eta_{\sigma,\sigma^{\prime}}$ gives the spin
degrees of freedom, which takes value of 2 for transverse spin and 1 for
longitudinal and charge susceptibility. The model involves three Green's
functions, the bare $G_0$, the dressed $G$ given by Dyson's equation
$G^{-1}=G_0^{-1}-\Sigma$, and an internal Green's function $G_{int}$ which
will be described further below.  A number of different variants of the GW
scheme can be constructed, depending on the specific Green's function used
in evaluating the $G$ and the $W$ in the convolution integral of
Eq.~1.\cite{gw} Using the bare $G_0$ in both $G$ and $W$ in the so-called
`$G_0W_0$' scheme corresponds to lowest order perturbation theory. Using
the dressed $G$ in both $G$ and $W$ (i.e. the $GW$ scheme) leads to fully
renormalized propagator and interaction corresponding to an infinite
resummation of diagrams. However, this is still not the exact self-energy
because of the missing vertex corrections. In fact, $GW$ scheme often
gives worse results than $G_0W_0$ version when the vertex corrections are
omitted. Bearing all this in mind, our approach is an intermediate one, in
that it is based on the convolution of an intermediate coupling Green's
function and interaction. In this sprit, we first calculate the
self-energy of Eq.~1 by using a parameterized $G=G_{int}(Z)$, and then
calculate $W_{int}$ exactly based on this $G_{int}$, and determine the
renormalization parameter $Z$ self-consistently.

More specifically, we write $G_0$ in terms of {\it the unrenormalized LDA
dispersion}. Note that our tight-binding hopping parameters are not free
parameters, but are the best representation of the first-principles LDA
dispersion. All renormalizations, giving rise to the experimental results,
are embedded in the computed $\Sigma$. $W$ is the sum of the RPA spin plus
charge susceptibilities calculated using $G_{int}$ rather than $G_0$. The
key lies in the choice of $G_{int}$.  Our strategy is to construct the
best one parameter model for $G_{int}(Z)$ with $Z$ chosen to minimize
$G-G_{int}$. [$G_{int}=G$ of course yields the full $GW$.] To motivate our
choice, we recall that the main effect of $\Sigma$ at low energies is to
renormalize the dispersion from the LDA values to those observed in
experiments (e.g., ARPES).  This renormalization, which amounts to a
factor of 2-3, is relatively modest in that the mass does not diverge, and
depends weakly on $k$.\cite{Arun3} That is, approximately
\begin{equation}\label{eq:1}
\epsilon_{{\bf k},ARPES}=Z\epsilon_{{\bf k},LDA}.
\end{equation}
Hence, we choose perhaps the simplest $G_{int}$ which reproduces this
dispersion renormalization, $G_{int}^{-1}=G_0^{-1}-\Sigma_{int}$, with
$\Sigma_{int} =(1-Z^{-1})\omega$, so that
\begin{equation}\label{eq:2}
G_{int}({\bf k},\omega )={Z\over\omega -Z\epsilon_{{\bf k},LDA}+i\delta}.
\end{equation}
The above expression refers to the paramagnetic phase, but the extension
to a magnetically ordered phase is straightforward where $G$, $\chi$, and
$\Sigma$ become $(2\times 2)$ tensors for the $(\pi ,\pi )$
antiferromagnetic order.\cite{vignale,tanmoysw}

\begin{figure}[h]
\centering
\rotatebox{270}{\scalebox{0.5}{\includegraphics{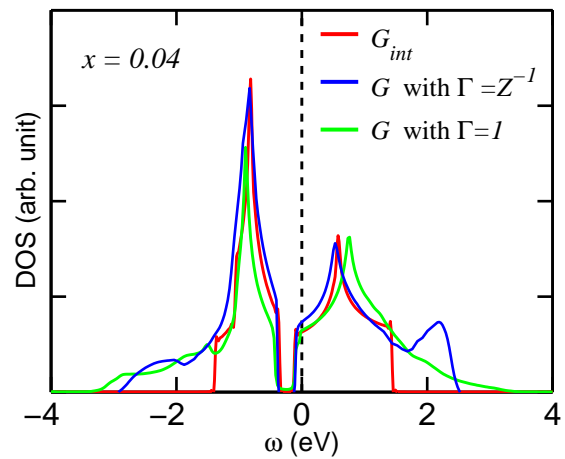}}}
\caption{DOS computed from $G_{int}$ is compared with the fully dressed
DOS with a vertex correction $\Gamma=1/Z$ and without the vertex
correction (i.e. $\Gamma=1$) at a representative doping of $x$=0.04 as
discussed in the text.}
\label{dos}
\end{figure}
\begin{figure*}[top]
%\hspace{-1.2in}
\centering
{\rotatebox{270}{\scalebox{0.7}{\includegraphics{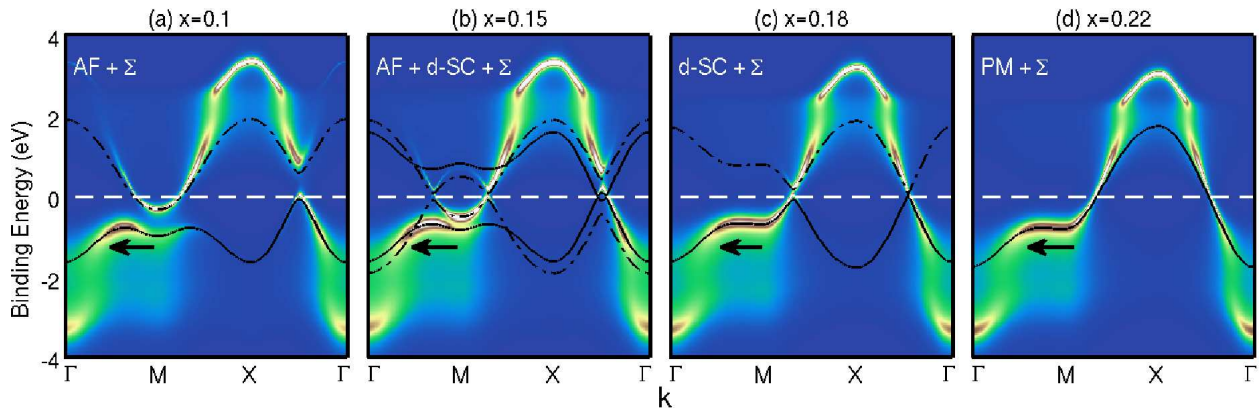}}}}
\caption{Spectral intensity plots along the high symmetry lines for
antiferromagnetic (AF) ($x=0.10$), AF+$d-$wave superconducting (SC)
($x=0.15$), $d-$SC ($x=0.18)$, and the paramagnetic (PM) ($x=0.22$)
states. Full self-energy $\Sigma$ is included in all cases.
Antiferromagnetic order parameters are computed
self-consistently.\cite{tanmoysw} Self-energy is computed including the
antiferromagnetic gap, but does not incorporate the low energy
superconducting gap. Black lines depict the dispersions that enter into
$G_{int}$, but the corresponding spectral weight of the band is not shown
for simplicity\cite{kusko}. An artificially large value of superconducting
gap ($\Delta=30$~meV) is used so that the effects of superconductivity can
be visible on the energy scale of the figure.\cite{tanmoyprb} Arrows mark
the start of the high-energy kink or the waterfall in the spectrum below
the Fermi energy.}
\label{disp}
\end{figure*}
Self-consistency is obtained by choosing $Z$ such that the low energy
dispersion is the same for $G$ and $G_{int}$.  This is illustrated in
Figs.~1 and~2. Fig.~1 compares the density-of-states (DOS) associated
with $G_{int}$ (red line) and $G$, either with (blue line) or without
(green) a vertex correction in the antiferromagnetic state. As discussed
further below, $G$ and $G_{int}-$dressed DOSs clearly match well with each other
in the low energy regime where both spectra show the spin density wave
dressed upper and lower magnetic bands, consistent with our earlier
mean-field results\cite{kusko,tanmoyprb,tanmoyprl}. At high energies,
however, $G_{int}$ fails (by construction) to reproduce the incoherent
hump features associated with precursors to the upper and lower Hubbard
bands.

Fig.~2 compares the spectral weights, $A=-Im(G)/\pi$, with the
corresponding dispersions $\epsilon_{{\bf k},int}$ (black lines) that
enter into $G_{int}$. As noted earlier, the latter provide a very good fit
to the low energy coherent features for the entire doping range in
electron doped Nd$_{2-x}$Ce$_x$CuO$_4$ (NCCO). Self-consistency ensures
that $G_{int}$ provides the best approximation to the full $G$. As an
added benefit, $Z$ in general renormalizes $\epsilon_{{\bf k},int}$ to
values close to those found in our earlier mean-field
studies\cite{kusko,tanmoyprb,tanmoyprl}. The obtained
self-consistent values of $Z$ decrease almost linearly with doping, as
seen in ARPES\cite{sahrakorpiprb}. In this way, the QP-GW model reproduces
the results of our earlier mean-field calculations in the low energy
region, while revealing new physics at higher energies [e.g., the
waterfall effect].

We now comment on some applications of the present QP-GW model. The
waterfall physics seen in ARPES spectra of the cuprates is a direct
consequence of the self-energy correction, which introduces a peak in
scattering at intermediate energies below as well as above the Fermi level
as seen in Fig.~2.\cite{susmita} This scattering splits the spectrum into
a low-energy coherent part and a high-energy incoherent region. While the
near-Fermi-level dispersion changes substantially as the magnetic and the
superconducting phases evolve with doping, the overall energy regime of
the waterfall phenomenon remains fairly doping independent (marked by
arrows in Fig.~2), consistent with experiments.\cite{graf} In the
pseudogap region ($x=0.10$, Fig.~2(a)), the resulting `four band'-like
structure (two magnetic bands and the two Hubbard bands) agrees well with
cluster\cite{grober} and quantum Monte Carlo calculations\cite{jarrell}.
Near optimal doping $d-$wave superconductivity coexists with the
antiferromagnetic state in a uniform phase\cite{tanmoyprb,tanmoyprl}
resulting in further splitting of the coherent bands as seen in Fig.~2(b).
The coherent bands approach the Fermi level with increasing spectral
weight as the pseudogap collapses at a quantum critical doping near
$x=0.17$ in both the electron and hole doped case\cite{tanmoy2gap}. On the
other hand, the Hubbard bands move towards higher energy as the doping
increases and the spectral weight associated with these bands decreases,
consistent with optical spectra.\cite{uchida,onose}

\begin{figure}[h]
\centering
\rotatebox{270}{\scalebox{0.5}{\includegraphics{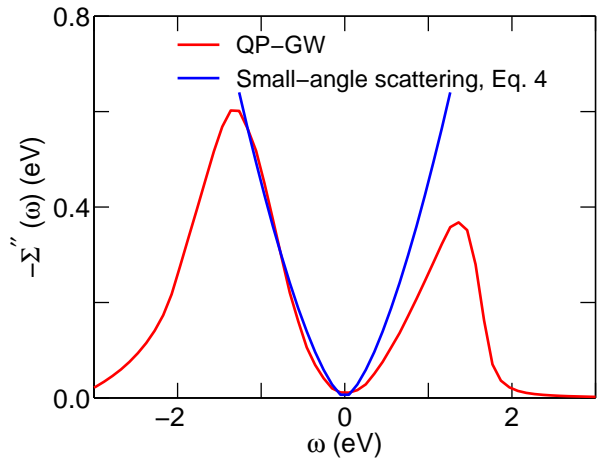}}}
\caption{Imaginary part of the self-energy for NCCO (red line) is compared
with the form used in Eq.~\ref{scattering}. Note that the model small
angle scattering formula was only applied to filled states, $\omega <0$. }
\label{dos}
\end{figure}
Notably, the lifetime broadening of the quasiparticle states originates
from magnetic scattering in the QP-GW model.  This broadening has a
non-Fermi liquid form with a significant linear-in-T component,
particularly near the Van Hove singularity (VHS).\cite{markiewater}
Furthermore, in describing broadening of ARPES features in the
superconducting state in our earlier mean-field model, we
phenomenologically introduced an elastic small angle scattering
contribution of similar non-Fermi-liquid form\cite{tanmoyprb,hlubina,dahm}
\begin{equation}\label{scattering}
\Sigma^{\prime\prime}(\omega)= sgn(\omega)C_0\left[1+\left(\frac{\omega}{\omega_0}\right)^p\right].
\end{equation}
Here $C_0=100$~meV and $\omega_0=1.6$~eV are determined from a fit to the
ARPES energy distribution curves (EDCs). The exponent $p$ is of physical
significance in determining quasiparticle character\cite{Vidhyadhiraja}.
We found that $p=3/2$\cite{tanmoyprb,hlubina,dahm} applies for electrons as well as
hole doped cuprates in reproducing the ARPES spectra and quasiparticle
interference (QPI) pattern seen in scanning tunneling spectroscopy. This
scattering is particularly important in that it allows a finite spectral
weight near the Fermi level even in the presence of a pseudogap, thereby
revealing the leading edge superconducting gap at all momenta\cite{tanmoyprb,tanmoy2gap}. In Fig.~3
we show that the imaginary part of the self-energy computed from Eq.~1
above reproduces this phenomenological form very well in the low-energy
region, both in magnitude and $T$-dependence, indicating that magnon
scattering underlies the anomalous exponent $p=3/2$.

We emphasize that in our QP-GW model the bare dispersion is taken directly
from LDA, and the self-consistently determined $Z$ renormalizes this into
a dispersion which matches the bands seen in ARPES experiments. Hence, the
model reproduces our earlier mean-field results, but with {\it fewer
parameters}, since the dispersion is calculated self-consistently
rather than being derived from experiment.

As noted above, the high energy features are absent in $G_{int}$. This is
also clear from Eq.~\ref{eq:2}: If we integrate the spectral weight
$A_{int}({\bf k},\omega)$ over all frequencies we get $Z$, not $1$, so
that $G_{int}$ accounts for only the coherent part of $G$. The incoherent
part of weight, $1-Z$, is thus not accounted for. Clearly, this could be
done straightforwardly by including a pair of broadened Lorentzians, but
this will add additional parameters in the computation of the self-energy,
not to mention associated vertex corrections. Therefore, we have chosen to
first explore the QP-GW model without these complications. This is also
the reason for calling our approach as QP-GW because it focuses on the QP
part of the spectrum in evaluating $G_{int}$.

In order to better understand the susceptibility $\chi_{int}$, a
comparison with our earlier mean-field result, $\chi_{MF}$ is instructive.
Since $G_{int}$ differs from $G_{MF}$ only by an overall multiplicative
factor of $Z$, $\chi_{int}$ differs from $\chi_{MF}$ by a factor of $Z^2$.
This has important consequences. Fermi liquid theory requires that both
the dispersion and the spectral weight be renormalized by interactions,
and for a weakly $k$-dependent self energy, as in the present case,
$Z_{disp}\simeq Z_{\omega}$. This is true for $G_{int}$ and $G$, but
$Z_{\omega}=1$ for $G_{MF}$, which causes mean-field theory to
overestimate the tendency towards instability.  As the bandwidth decreases
($Z\rightarrow 0$), the density of states and susceptibility must increase
as $1/Z$, in order to keep the total electron number fixed since there are
no incoherent states in mean-field theory. Instability is controlled by
the Stoner factor, $U\chi_0({\bf q},\omega = 0)=1$.  Since $\chi_0({\bf
q},0,MF)=\chi_0({\bf q},0,LDA)/Z$, a small $Z$ enhances the probability of
instability. On the other hand, for $G$ or $G_{int}$ only the low-energy
quasiparticle degrees of freedom contribute to the instability, as
reflected in $\chi_0({\bf q},0,int)=Z\chi_0({\bf q},0,LDA)$.  Hence, large
fluctuations (small $Z$) actually reduce the probability of condensing
into any one mode. Equivalently, if we rewrite the Stoner factor in terms
of the LDA susceptibility, it becomes $U_{eff}\chi_{0,LDA}=1$, with
$U_{eff}=ZU$. Thus $t$ and $U$ should both be renormalized by factors of
$Z$, leaving $t/U$ invariant.  In our mean-field treatment, we had to
assume that the effective $U$ was doping-dependent, and this
$Z$-correction accounts for part of that doping dependence. .

It should be noted that an accurate calculation of the susceptibility and
the resulting self-energy is fairly computer intensive as it involves a
three-dimensional integral ($k_x,k_y,\omega$) for $\chi_0$ and a similar
three-dimensional integral over $\chi$'s for $\Sigma$. Fortunately, we
find that $\Sigma$ has only a weak $k$-dependence, so we need to calculate
it only over a few $k$-points and use the average. Clearly, this is not a
limitation of the model, and the full $k$-dependence could be calculated.
However, this would make accurate self-consistent calculations
substantially more time intensive. We have explored the use of a vertex
correction for $\Sigma$, but the results are not too sensitive. We
have typically taken $\Gamma =1/Z$, which puts somewhat greater weight
into the incoherent bands as seen by comparing blue and green curves at
higher energies in Fig.~2.

The present scheme can straightforwardly incorporate the full
$k$-dependence of the susceptibility based on a realistic material
specific band structure.\cite{Gzm1} This is important for delineating the
nature of competing ordered phases, which are different for electron and
hole doped cuprates. Moreover, our self-energy provides a tangible basis
for going beyond the conventional LDA-based framework for realistic
modeling of various highly resolved spectroscopies, providing more
discriminating tests of theoretical models. In addition to the ARPES
spectra discussed above, a note should also be made in this connection of
the optical spectroscopy,\cite{tanmoysw} STM \cite{tanmoytwogap,Jouko,joukoprb}, RIXS\cite{susmitarixs,RIXS}, x-ray absortion spectroscopy (XAS)\cite{towfiq} and other inelastic light scattering
spectroscopies\cite{Tanaka,Huotari,Barbiellini,bansil} to help piece
together a robust understanding of the nature of electronic states in the
cuprates and their evolution with doping.

In summary, we have shown that our intermediate coupling model of
self-energy, which is based on the spin-wave dressing of the
quasiparticles, can explain many anomalous features of the cuprates. At
low energies, the model reproduces our mean field results for the coherent
bands in  ARPES,\cite{susmita} optical,\cite{tanmoysw} and RIXS,\cite{susmitarixs} with self-energy corrections renormalizing the
large widths of the LDA dispersions.  At high energies, we obtain the
waterfall features which represent a splitting off of the incoherent
bands, precursors of the Mott gaps seen in ARPES and optical studies.  In
the underdoped regime, the coherent in-gap bands reproduce both the
four-band behavior seen in quantum cluster calculations and the magnetic
gap collapse found in the mean-field calculations and a variety of
experiments. These results clearly suggest that the cuprates can be
understood within the intermediate coupling regime with an effective $U$
value substantially smaller than twice the bandwidth.

\begin{acknowledgments}
This work is supported by the US Department of Energy, Office of Science,
Basic Energy Sciences contract DE-FG02-07ER46352, and benefited from the
allocation of supercomputer time at NERSC, Northeastern University's
Advanced Scientific Computation Center (ASCC).  RSM's work was partially
funded by the Marie Curie Grant PIIF-GA-2008-220790 SOQCS.
%\end{acknowledgments}
\end{acknowledgments}
%\section*{The Bibliography}
\bibliographystyle{elsarticle-num}

\begin{thebibliography}{99}
\bibitem{foot0} The self energy is calculated in Matsubara space and
continued analytically as $i\omega_n\rightarrow\omega +i\delta$ where $\delta$ is a positive infinitesimal.
\end{thebibliography}

\end{document}